*Article*

# Improving Emergency Training for Earthquakes through Immersive Virtual Environments and Anxiety Tests: A Case Study

Mohammad Sadra Rajabi [1,2,*], Hosein Taghaddos [2,*] and Seyed Mehdi Zahrai [2,3]

[1] Department of Industrial and Systems Engineering, Virginia Tech, Blacksburg, VA 24061, USA
[2] School of Civil Engineering, College of Engineering, University of Tehran, Tehran 1417935840, Iran; mzahrai@ut.ac.ir
[3] Civil Engineering Department, University of Ottawa, ON K1N 6N5, Canada
* Correspondence: rajabi@vt.edu (M.S.R.); htaghaddos@ut.ac.ir (H.T.)

**Abstract:** Because of the occurrence of severe and large magnitude earthquakes each year, earthquake-prone countries suffer considerable financial damages and loss of life. Teaching essential safety measures will lead to a generation that can perform basic procedures during an earthquake, which is an essential and effective solution in preventing the loss of life in this natural disaster. In recent years, Virtual Reality (VR) technology has been a tool used to educate people on safety matters. This paper evaluates the effect of education and premonition on the incorrect decision-making of residents under the stressful conditions of an earthquake. For this purpose, a virtual model has been designed and modeled based on a proposed classroom in a school in the city of Tehran to simulate a virtual learning experience. In contrast, the classroom represents a realistic method of learning. Accordingly, each educational scenario, presented in reality and the virtual model, respectively, was conducted on a statistical sample of 20 students within the range of 20 to 25 years of age. Among the mentioned sample, the first group of 10 students was taught safety measures in a physical classroom. The second group of 10 students participated in a virtual classroom. Evaluation tests on safety measures against earthquakes were distributed after two weeks. Two self-reporting tests of Depression, Anxiety, Stress Scale (DASS) and Beck Anxiety Inventory (BAI) tests were assigned to the second group to evaluate the effect of foresight under two different scenarios. The results indicate that teaching through VR technology yields a higher performance level than the in-person education approach. Additionally, the ability to detect earthquakes ahead is an influential factor in controlling anxiety and determining the right decisions should the event occur.

## 1. Introduction

To mitigate the casualties and financial damages caused by earthquakes, it is vital to design and construct buildings resistant to natural disasters. Moreover, timely reactions and proper evacuation practices following an earthquake significantly reduce earthquake-related losses [1,2]. Thus, earthquake-prone countries have issued a list of suggested measures for dealing with dangers during and after an earthquake [3]. Seminars, maneuvers, posters, lectures, and instructive movies have all been offered as options for conveying these standards to the public. However, these informative techniques lack depth and emotion and cannot fully represent the message to viewers. As a result, the effectiveness of these teaching initiatives frequently falls below expectations [4].

Technological advances have been addressing engineering and construction challenges in recent years, particularly in construction safety and facility management [5–11]. Prior studies used different technologies and tools, including social media platforms [12],

Immersive Virtual Reality (IVR), and Serious Games (SG) to educate people regarding safety matters [13,14]. IVR is a technique that involves a person in a computer-generated virtual world actively. More realistic risks and threats may be simulated and presented to participants using this technology, allowing for the creation of desirable scenarios in educational settings. SG, otherwise known as educational games, are a type of video game in which one of the primary purposes is to educate the player [14]. Using VR technology, serious games can effectively aid in developing stated educational goals. These technologies have been widely used for educational purposes such as surgery training [15], teaching the repair and maintenance of high-voltage power lines [16], and setting pedestrian safety guidelines [17]. They are also used for engineering purposes, such as programming heavy mobile cranes for construction sites [18]. However, these tools have not been significantly applied to teaching earthquake safety precautions before, during, and after an earthquake [19].

This study aims to demonstrate IVR's capabilities during earthquake-related safety training and to evaluate how educating, foretelling seismic activity, and simulating earthquakes in IVR affect people's reactions to earthquakes. From most people's perspectives, the real experience of an earthquake is enormously increasing stress, anxiety, and situational awareness, which is a lack of traditional in-person educational approaches. Therefore, this study intends to implement VR technology to provide a near-real experience of earthquakes and illustrates the application of VR in earthquake safety training. First, IVR, DASS, and BAI are thoroughly examined. The design and development of IVR are then incorporated into a case study based on a proposed high school in Tehran. Participants are divided into two groups, and the designed experiment is executed. Finally, the preliminary findings and their implications for earthquake safety guidelines and practices are presented.

## 2. Literature Review

In the past, schools and offices have taken the initiative to provide earthquake-related safety training, usually in the form of a brief informational session led by a professional. Prior to the 2011 study [20], unofficial assessments were conducted in educational settings. Kirikaya and his team explored the impact of such teachings on students' understanding of earthquakes and related phenomena. According to the findings, almost half of the students are unaware that they reside in a seismically active area.

California's San Francisco gulf region is one of those seismically active areas. In a study of San Francisco's gulf region by Simpson [21], residents were taught basic survival and emergency response skills in educational programs for assessment. In these programs, CPR, search and rescue techniques, and fire extinguishing lessons were given to locals to prevent earthquake-related casualties in the region. According to this research study, more than a hundred of these educational programs are currently available. Many of them are provided in annual courses that teach and evaluate the aforementioned subjects.

### 2.1. IVR for Earthquake Safety Training

The outstanding capabilities of VR technology have motivated researchers to employ it for immediate reactions before, during, and after an earthquake. For example, Gong et al. [22] utilized VR to simulate a dormitory exposed to an earthquake in a three-dimensional environment, utilizing a monitor in front of their faces. Such an immersive simulation model enabled the participants to experience an earthquake and communicate with a virtual environment through Kinect, a line of motion sensing input devices, and SIGVerse (i.e., A cloud-based VR platform). The results indicated that powerful earthquakes can be simulated successfully within VR technology and can be beneficial for earthquake exercises or drills [22].

In another educational approach, Li et al. [23] utilized VR technology to demonstrate how to survive and await rescue teams while remaining inside a building during an earthquake. The study aimed to show that safety training programs against earthquakes could be presented through VR. VR allows users to learn the necessary skills for staying alert during an earthquake realistically by submerging in a virtual environment.

Similarly, Feng et al. [19] displayed how serious games in VR could be implemented to teach an evacuation of a building during emergencies (e.g., fires and earthquakes). For this purpose, a conceptual framework for effectively designing and implementing VR has been created by systematically reviewing technical literature. In another similar study, Lovreglio et al. [24] evaluated the theoretical advantages and disadvantages of using serious games based on VR for assessing people's behavior during an earthquake. The building residents should follow regulations to protect themselves during the stressful conditions of an earthquake. Precise and repetitive training courses are required to enforce these rules. Consequently, the authors discussed the main elements of designing and developing a VR-based educational system based on serious games for the aforementioned conceptual framework. Additionally, Fang et al. [25] have cited that strengthening timely and intelligent reactions during an earthquake helps decrease the mortality rate and structural damage in another study on Auckland Hospital.

Concerning government-related responses, The New Zealand Civil Defense Agency has provided a safety guideline for earthquakes, including 32 measures to enact during and after the time of an earthquake [26]. It includes the appropriate responses across a broad spectrum of scenarios, such as instructions for home, the workplace, places away from home, coastal areas, mountainous regions, vehicles, and locations with domestic or farm animals. In contrast, research conducted by Fang et al. mainly concentrated on indoor environments to evaluate behavioral responses modeled within the VR environment [19,25].

Fang et al. [27] have also performed a research study on educating earthquake-related conduct to [school-age] children. The researchers in this study claimed that children are very vulnerable and should be educated to create a more knowledgeable society on the subject of earthquakes. Therefore, to successfully prepare children against earthquakes, an effective educational tool must be considered. A VR-based educational system for serious gaming was developed in this research study based on Problem-Based Learning (PBL). The educational system's structure consisted of three mechanisms: previous education, instant feedback, and evaluation. These mechanisms were deployed after a completed game to assess how to increase a user's level of proficiency. The results indicated that training children in a VR-based serious educational game were the most effective way of improving their self-efficiency due to the performance analysis function within the education system [27].

Various studies concluded that ensuring safety throughout an earthquake evacuation process is vital to decreasing the number of deaths and damages sustained to buildings. Despite scientists attempting different ways of predicting earthquakes, Liang et al. [28] pointed out the impossibility of determining the precise timing and frequency of its occurrence. To accomplish this, the researchers designed a primary sample VR-based system. This study claims that the developed VR system can evaluate participants' behavior by simulating the structural and non-structural damages from earthquakes and exercising different emergency scenarios. The provided framework can also contribute to feedback and lay a foundation for improving the necessary behaviors for an earthquake [28].

In another study involving behavioral analysis, Fang et al. [29] introduced a virtual experiment that evaluated the decisions of a building's residents during and after an earthquake. This VR experiment involved 83 participants (i.e., staff members and patients from Auckland Hospital) positioned on a shaking table to simulate the earthquake vibrations. Then, the participants' cognitive abilities were assessed based on verbal protocol analysis (i.e., expressing thoughts orally) to understand the decision-making process. Through this analysis, basic reasoning and other mental processes are transparent for observation [29–33].

Two Auckland Hospital staff members assisted by giving instructions throughout the VR simulation to test compliance when authoritative figures were present. Most of the participants relinquished their autonomy and acted according to the staff members' decisions. However, 13 percent of the group responded differently by ignoring the staff members and trusting their instincts. These results generally show that people will follow instructions, especially those in a position of power [29].

*2.2. Depression, Anxiety, and Stress Scale (DASS)*

The most common psychological disorders in modern times are depression, anxiety, and stress. Depression is a psychological disorder that causes hopelessness and loss of interest in formerly enjoyable activities. Most people have a tendency to (at times) feel depressed and disconsolate. After all, it is the body's natural reaction to problems in life, resulting in a lack of enthusiasm for relevant people and interests. The person is often diagnosed with depression when this sensation of severe despondency, a sense of hopelessness, and disheartenment lasts more than a few days or weeks [34,35]. A similar process occurs when anxiety emerges, and a person can be diagnosed with it when the condition interferes with his or her daily life. It is an intense feeling of worry, fear, and uncertainty from an unknown source. Although anxiety is sometimes natural, chronic and severe anxiety can be problematic and unusual. The average individual with anxiety can be constantly afraid, troubled, and stressed [35].

As for stress, its symptoms are related to anxiety, and it feels like a sensation of pressure. It can be overwhelming when a person is not adjusted to handling certain stress levels. When someone undergoes stress, his or her breathing rate, and heartbeat increase, resulting in more energy [36]. Stress is common and even beneficial in smaller quantities because of the motivation it gives. Some individuals believe that stress helps them and that life loses its meaning without it [37]. However, stress might have negative consequences when it impacts a person's daily life and overpowers their ability to function properly. This is called bad or ablative (decreasing) stress [34,38,39]. These three psychological disorders (i.e., depression, anxiety, and stress) are interconnected, which is why psychologists will examine them together when formulating a diagnosis [39].

Diagnosis is the prerequisite for treating psychological disorders. However, various approaches to detecting psychological disorders exist. The most reliable way is to visit a psychiatrist or psychologist. Nevertheless, some people will feel that their problems cannot be severe and will search for a more convenient way of diagnosis. Using accredited psychological tests is an easier way of verifying if someone has a psychological disorder without consulting a medical professional. One of the best-accredited tests, designed by Lavibound [40], was named the DASS. This test is proved to be promising because of its accuracy during many trials [34,35,38–42].

The DASS has two different forms. The main form of DASS has 42 questions that measure each psychological component every 14 questions, while the shortened form contains 21 questions that measure each psychological factor every seven questions [40]. In a study by Tellegen et al. [43], the team of researchers evaluated the DASS for the factors of depression, anxiety, and stress. The results of this study showed that these three factors considered 68% of the variance on all scales. The special values for stress, depression, and anxiety were 9.07, 2.89, and 1.23, respectively, and the alpha coefficients for these factors were 0.97, 0.92, and 0.95, respectively. When Tellegen et al. [43] calculated the correlation between all of the factors, it was revealed that there was a correlation coefficient of 0.48 between depression and stress, 0.53 between anxiety and stress, and 0.28 between anxiety and depression.

In 2005, Henry and Crawford [44] studied the shortened form of the DASS for the validity of its structure and accounted for the factors of depression, anxiety, and stress. Henry and Crawford reported the final coefficients for all of the factors, which were equal to 0.88, 0.82, 0.9, and 0.93, respectively. Because of the relationships between depression, anxiety, and stress, and especially the relationship between anxiety and depression, several studies in Iran have chosen to use the shortened form of DASS.

In 2007, Samani and Jokar [45] also evaluated the validity and reliability of the DASS. They discovered that the reliability of depression, anxiety, and stress were equivalent to 0.8, 0.76, and 0.77, respectively, and that the Cronbach alpha for those same variables was equivalent to 0.81, 0.74, and 0.78, respectively. In the same year, Asghari et al. [46] conducted another research study involving the same subject but on non-clinical samples. The goal was to determine the reliability and validity of the DASS within the greater context of society. They performed the Beck Anxiety Inventory (BAI) and the Four-Dimensional Symptom Questionnaire (4DSQ) on a sample of 420 adults to test the effectiveness of the depression, anxiety, and stress scale.

The evaluation of internal similarity and test-retest coefficients confirmed the validity of the DASS. The exploratory factor analysis showed that 14 phrases were attributed to stress, while depression and anxiety had two phrases attributed to them each. The two phrases for depression and anxiety greatly impacted all three psychological aspects. When these four phrases were removed, the analysis confirmed the DASS structure's exploratory factor and its exceptional operation. Additionally, the correlation calculation method of the sample scores in the Beck depression scale, and the four orders anxiety questionnaire confirmed the reliability of the DASS structure. The concurrent credibility of the DASS was confirmed by comparing the scores of a subsidiary sample taken from the general population (315 people) with a counterpart group of patients who had psychological disorders (130 people). Based on the results of this research study, it can be concluded that the depression, anxiety, and stress scale meets the required conditions for psychological research or clinical settings in Iran [46].

In the DASS evaluation test, all three factors are witnessed in a person. The individual percentages of depression, anxiety, and stress can be observed in the test results [42]. The standardized cutoffs in this scale for anxiety are 0–16 as natural, 17–21 as weak, 22–33 as moderate, 34-45 as Severe, and equal or greater than 46 as extremely severe levels.

*2.3. Beck Anxiety Inventory (BAI)*

Assessing anxiety symptoms is essential for diagnosing and treating anxiety. However, numerous scales propose different perspectives, leading to a lack of standardization [47–49]. These scales most likely contain problems regarding their conception and psychological properties [50,51]. Beck et al. [52] considered these problems and introduced the BAI, which precisely measures the intensity of patients' anxiety symptoms.

The BAI is a self-reporting questionnaire that measures the intensity level of anxiety in teenagers and adults. Conducted studies prove that this questionnaire is highly credible. Its internal consistency coefficient (Alpha coefficient) is equal to 0.92, its reliability after one week using the test-retest method is equal to 0.75, and the correlation of its clauses varies from 0.30 to 0.76. Five correlations between content, concurrent, structure, diagnosis, and factorial are measured in this test, which all show the efficiency of BAI in measuring anxiety intensity [53]. The psychological components of this test have also been investigated. Gharaie et al. [54] reported its reliability coefficient through the test-retest method after two weeks. Kaviani and Mousavi [55] also assessed the psychological components of BAI in the Iranian population. They reported that the coefficients were almost 0.72, with the retest reliability coefficient equal to 0.83 after one month and the Cronbach alpha equal to 0.92.

Regarding its contents, the BAI has 21 questions with four options each. Each question describes one of the common anxiety symptoms (mental, physiological, and fear). Based on the participant's response, each answer will be ranked on a scale of 0 to 3. Therefore, the total score of this questionnaire ranges from 0 to 63. The standardized cutoffs in this scale are 0–7 as Minimal, 8–15 as Mild, 16–25 as Moderate, and 26–63 as Severe levels.

*2.4. Research Gap*

Researchers have evaluated human behavior and thought processes in recent studies by simulating earthquakes in virtual public areas [19,24,27,29]. According to Feng et al. [29], comprehending decision-making processes can explain people's reactions to different conditions in an earthquake. This information can assist in developing new guidelines and proper evacuation methods after an earthquake. Furthermore, it can provide essential data for enacting earthquake-related safety policies and improving safety training on this natural disaster. Although these efforts have advanced the quality of education on safety procedures, to the best of the authors' knowledge, no research has been conducted to find how certain psychological factors are affected by earthquakes using a virtual environment. This study aims to thoroughly research this subject and close the research gap using self-reporting psychological tests. It also discusses the effect of previous alerts for the occurrence of the earthquake on anxiety and the performance of people's decision-making. Finally, it evaluates whether or not VR has improved the quality of people's educational training.

## 3. Research Methodology

A VR model of a classroom in Tehran was constructed in Unity software at the beginning of this study, along with the virtual educational scenarios in a virtual environment. After the virtual education model was completed, the real-life educational model was constructed precisely based on virtual education due to the suggested statistical sample of two types of education, virtual and real. Finally, data from the DASS, BAI, and assessment tests for earthquake safety information were collected and evaluated. In Figure 1, the study procedure has been depicted.

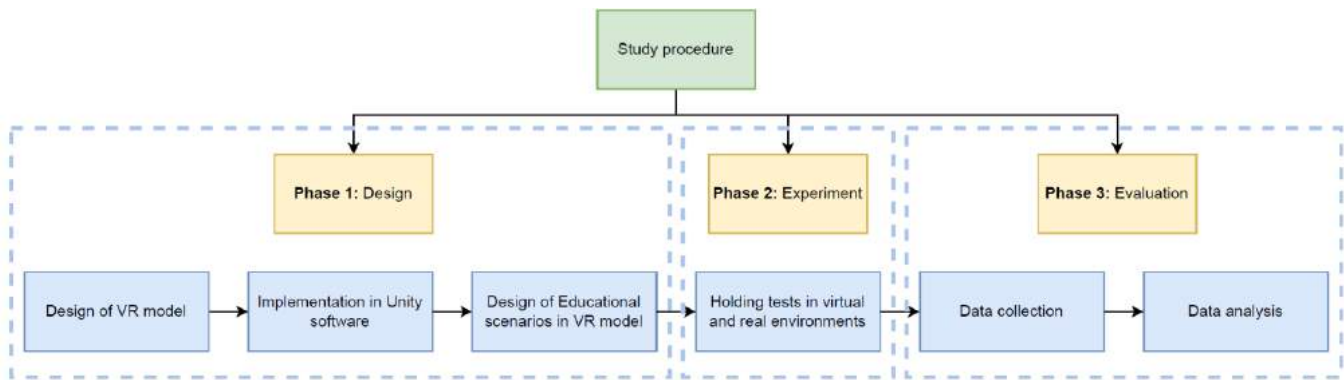

**Figure 1.** Study procedure.

### 3.1. Experimental Design

For the purposes of this research, the Oculus Rift, a type of VR headset, was used to display the VR environments. The user will have access to much more portable conditions when using this device. A VR model of a hypothetical class in Tehran was modeled using three-dimensional unity software. Unity is well known for its physics software and its animation options, which have been used to create realistic scenarios and life-like environments. To conduct a case study on students, different elements, parts, and assets have to be included within the modeled classroom environment on the Unity software. To define Unity assets, they are cases that can be used in games or projects. They may be created from a file composed of Unity software like a three-dimensional model, an audio file, an image file, or any other type of file that Unity supports.

To create a simulated environment similar to reality, objects, elements, and textures from the physical classroom environment in Tehran must be incorporated visually and graphically. Textures are video files or movies that exist in models or serious game elements and provide visual effects. The properties and capabilities of Unity must be closer to reality so the model can be more similar to how the classroom is in real life.

The floor is made from ceramic in the physical classroom environment, while half of the wall is composed of plaster and stone. Classrooms in Tehran commonly consist of these construction materials. Similarly, the model contains ceramic, plaster, and wall for a more realistic texture. The six benches used in the model are similar to the wooden benches used in Iranian schools. The body and frame of these benches are metal, and the seats and table of the bench are made of wood. Like Iranian benches, a metal box is under the bench surface. Some books, notebooks, pens, or pencils have been placed in front of the students. Each student's backpack has been modeled beside the bench or above the ground. Figure 2 illustrates a typical real classroom located in Tehran.

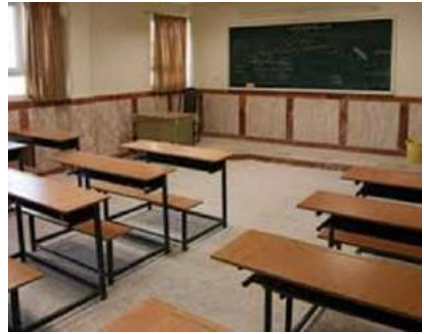

**Figure 2.** A typical real classroom in Tehran [56].

The students within the virtual model have supported the creation of a real classroom environment. All of them have been designed and modeled individually. As for their attributes, they have been given a blue uniform commonly seen in schools in Tehran. All aspects of their appearances are similar to those of real-life Iranian students. The student's pants, belts, and shoes have been designed based on current Iranian fashion trends. There are six students in the virtual classroom, including the user. The teacher is a dummy taller than the students and is equipped with gray hair, a professional suit, and a handbag. To create a natural classroom environment, these items have been hung on the wall: pictures of famous people in Iran, a map of the world, the periodic table, a whiteboard, a clock, and wallpaper made by a student commemorating the anniversary of the Iranian Revolution. Other elements in the model include three heaters on the classroom's wall, two bookshelves, garbage bins, two windows with a view of the city of Tehran, a teacher's desk and table, a fluorescent light bulb, and other light bulbs. There is also a short platform for teachers and presenters in front of the class, with two standing bookshelves on top of the mentioned platform.

Two different scenarios have been designed and implemented in the virtual classroom model. In the first scenario, the user awaits the teacher's entrance into the classroom while he or she is seated on a bench. Other students' voices can be heard throughout the classroom at this time, and they are located on other benches where notebooks, books, and pencils have been designated for them. The students model realistic movements, such as looking at the box under the table or looking around the classroom, so the virtual model can simulate reality. The user can freely observe any area of the classroom, such as the teacher's desk, benches, walls, windows, bookshelves, and other students, by moving around his or her head.

Underneath the virtual classroom, an engine room has been attached. At the end of the first scenario, a student falls into the engine room due to incorrect positioning. Figure 3 depicts several scenes from the first scenario.

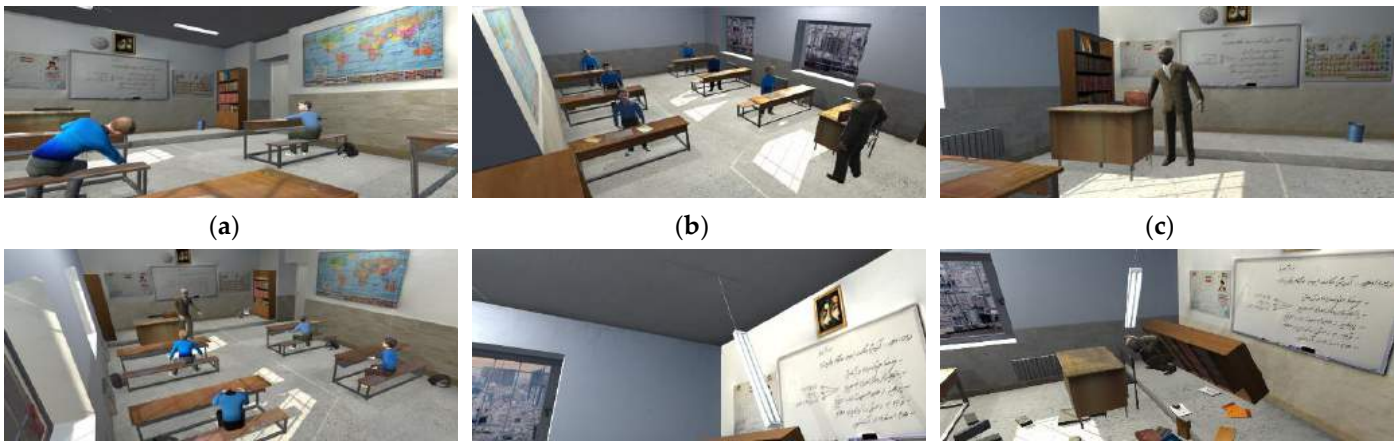

(**a**)            (**b**)            (**c**)

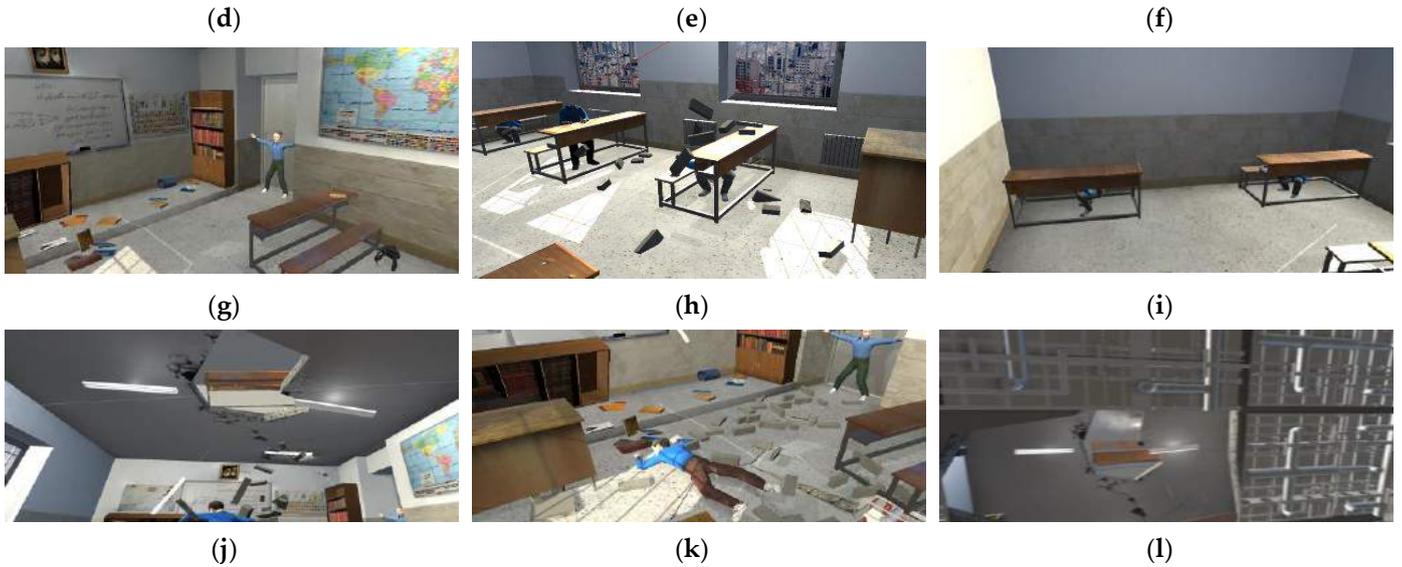

(**d**) | (**e**) | (**f**)
(**g**) | (**h**) | (**i**)
(**j**) | (**k**) | (**l**)

**Figure 3.** First educational scenario: Occurrence of an earthquake while safety measures were being taught virtually. (**a**) The user is waiting for the teacher; (**b**) Students are standing up to show respect to the teacher; (**c**) The teacher is explaining the earthquake safety measures; (**d**) a Shocking earthquake starts; (**e**,**f**) Falling class objects; (**g**) A student's incorrect positioning during an earthquake; (**h**,**i**) Correct positioning under the desk; (**j**,**k**) A student's incorrect positioning during an earthquake; (**l**) The designed engine room underneath the classroom.

After the user looks at the door, the teacher enters the classroom with a bag in his hand. The classroom representative orders the students to stand out of courtesy and respect for the teacher. The user stands, too, and observes the classroom from a higher point of view. During this time, the teacher walks toward his desk, places the bag down, and tells the students to sit.

Next, the teacher gives instructions for the lesson: "Hello. Class, you may be seated now. If you remember the lesson from last week, we discussed why an earthquake occurs. Today, we will talk about preparing for the occurrence of an earthquake while in school. All of you should not exit the door simultaneously. Instead, the best action plan is to locate the safest areas in the classroom and take shelter there until the building stops shaking. An example of a safe spot is underneath the benches. Hide under the benches and hold the legs tightly, so it stops vibrating. If you are in the library, workshop, or laboratory and cannot escape because of obstruction, keep a safe distance from the shelves and look for shelter. Keep away from the school building if you are in the schoolyard." These instructions are based on the guidelines of several organizations, namely the Centers for Disease Control and Prevention (CDC) [57] in the United States, the New Zealand Civil Defense Organization [26], and the United States Geological Survey (USGS) [58].

While the teacher is lecturing, the other students listen eagerly. The user can move his or her head around to look at the students or the teacher. For a more realistic experience, the teacher incorporates body language into his mannerisms. For example, the teacher moves both hands forward when he instructs the students to sit down and points to the section below the desks when he discusses where to hide in the classroom.

When the teacher reaches the last sentence, an earthquake erupts. The user will witness and feel the vibrations within the virtual environment. During this scene, the students will start screaming, and the teacher will inform them that they must take shelter immediately. After the initial shock subsides, some students will begin to defy orders and act differently. The simulation has been designed to illustrate the consequences of safe and unsafe actions. Because of this, the user can distinguish which behaviors are appropriate during an earthquake; this will be elaborated on in the following sections.

As the scene continues, other visual techniques are employed to add to the effect of realism. Several books fall at the front of the classroom, and the fluorescent light bulb above the teacher's head becomes detached from its wire. The teacher moves behind his desk, ordering students to stay calm and take shelter. Unfortunately, one of the bookshelves falls on top of the teacher, who screams out of pain.

Dust begins to cloud the atmosphere, and the students diversify in action. One of the students at the front of the classroom takes shelter under the door frame. Another student hides underneath a bench in the front row and puts his hands over his head. Immediately after the student does this, several chunks of the ceiling land on the bench. His quick and timely reactions show that the right choices can save a person's life.

As the building collapses, two students at the back of the classroom take shelter under their benches. A student on the user's right side, however, stands up from his bench and runs toward the front of the classroom. Plaster, soil, and bricks fall on top of the student and bury him, and the student stops moving. Various construction materials falling from the roof cause the floor to crack due to the weight of their impact. Consequently, the teacher warns students that the roof has collapsed and that students should not approach the area.

The user then stands up and proceeds to the middle of the classroom, where the damage to the floor is located. The floor's structure has lost its strength and can no longer support any weight, so the user falls through and lands in the engine room on the lower story of the building. The scene ends after this incident.

Once again, the simulation highlights the consequences of not taking shelter properly. The user is allowed to experiment with his or her choices and thus becomes acquainted with the right and wrong decisions during an earthquake. Through visual and auditory learning, the user learns how to react correctly, whether in a classroom, library, or laboratory.

The second scenario begins right after the first one ends. In this scenario, the user can observe safe and unsafe locations within the previous setting. Secure areas, such as the space underneath the benches or the corners in the back of the class, are marked by a green cube. Hazardous areas, such as the area in front of the bookshelves, the door frame, or the center of the classroom, are labeled with a red cube. Photos from the second scenario are presented in Figure 4. It should be mentioned that the user's perspective in this scenario is in front of the classroom and besides the teacher's desk.

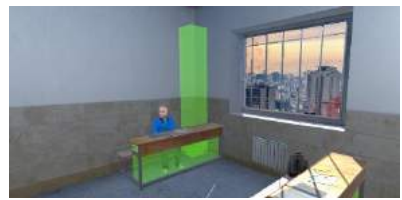
(**a**)

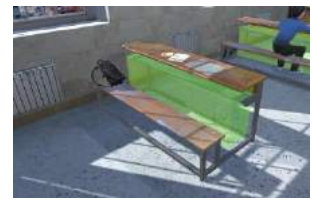
(**b**)

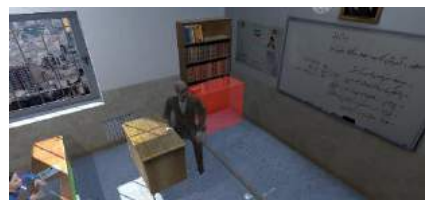
(**c**)

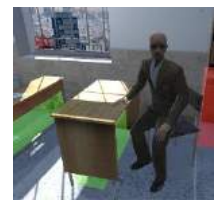
(**d**)

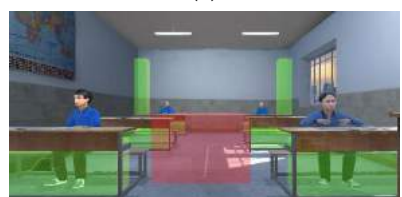
(**e**)

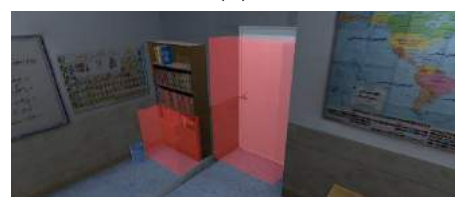
(**f**)

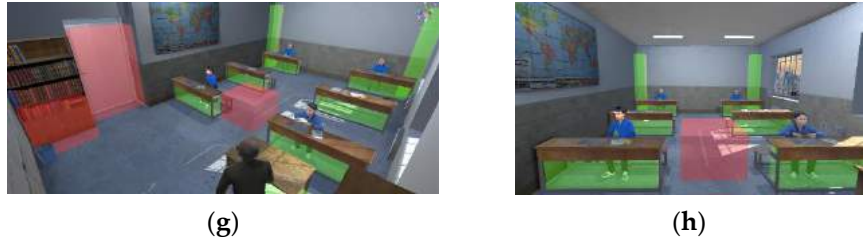

(**g**) (**h**)

**Figure 4.** Second Educational Scenario: Illustration of safe/unsafe areas in green/red color. (**a**) and (**b**) Corner of the class and under the bench are safe areas; (**c**) In front of the bookshelf is an unsafe area; (**d**) Under the table is a safe area for the teacher; (**e**) Center of the classroom and far from the columns is unsafe area; (**f**) In front of the bookshelf and entrance door are unsafe areas; (**g**) and (**h**) Overview of safe and unsafe areas in the classroom.

*3.2. Assessing VR on Safety against Earthquakes*

To evaluate how effective VR is on learning, a questionnaire was designed to ask the subjects about the learning objectives. With the assessments, the extent of this impact will be known. The questionnaire contained five questions, which are shown in Table 1.

**Table 1.** The questions of the assessment test on safety against an earthquake.

| Row | Questions |
| --- | --- |
| 1 | List the safe places for taking shelter in the classroom at the time of an earthquake. |
| 2 | In addition to previous cases, what other safe locations exist in the classroom during an earthquake? List all of them. |
| 3 | List the unsafe locations in the classroom during the Earthquake. |
| 4 | What type of safety measure should we take during an earthquake at a library, workshop, or laboratory? |
| 5 | What type of safety measures should we take in the school's yard during an earthquake? |

*3.3. Procedure*

The research study consisted of two groups of 10 male students, all aged 20 to 25. The first group was educated on earthquake-related safety procedures in a physical environment that adhered to COVID-19 protocols. To keep the physical and virtual models consistent, the topics corresponded to the first scenario in the VR classroom; each sentence spoken in the physical classroom followed what the teacher said in the simulation by 10 s or less. The first group returned two weeks after the initial training date to complete their assessment tests.

The second group was similar to the first group in terms of size, background, and age. The students were asked to finish the DASS and BAI questionnaires in an undisturbed environment. Next, they played through the two virtual scenarios discussed in the previous sections on an Oculus Rift. It should be noted that none of the students had used an Oculus Rift before and that for safety purposes, all COVID-19 safety protocols were implemented at this time. After completing both scenarios, they were asked to resubmit the DASS and BAI questionnaires. Just like the first group, the students in the second group completed their assessment tests two weeks after the initial date. Unlike the first group, however, the second group was divided into two subsections of 5 students. The first subsection was informed of the earthquake within the first scenario, while the second subsection was informed of the first scenario without an earthquake. The results of the assessment tests between the two subsections will be analyzed in the next section.

As mentioned before, the goal of the study is to compare the two groups against each other and to determine which method of education, in-person or virtual, produces a better learning curve. The first group attended a physical classroom with COVID-19 safety measures in effect, and they were taught how to navigate the dangers of an earthquake in the safest way possible. The in-person earthquake training session is shown in Figure 5. For the sake of consistency, the first group's curriculum mirrored the one in the VR model.

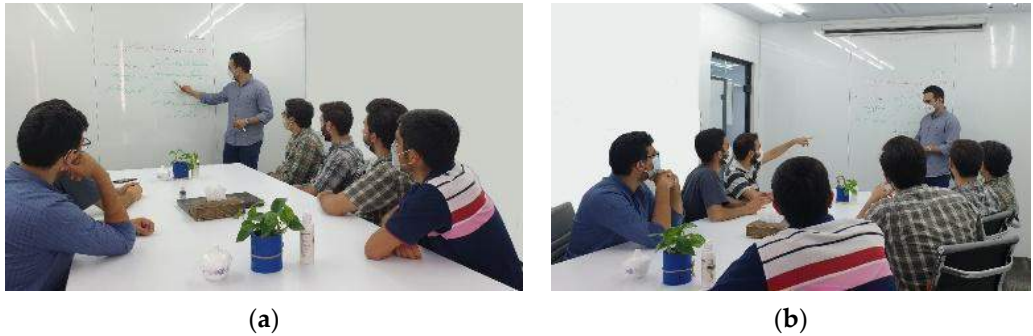

(a) (b)

**Figure 5. (a,b)** Earthquake training in a real environment.

The second group experienced the same training as the first group, except for some differences. The second group received training in a virtual classroom, and half of those students were not informed of an earthquake in the first scenario. All of the students were tested in one day.

Because VR is an immersive environment, the second group of participants was given a revolving chair to enhance their experience and allow their heads to move 360 degrees. The earthquake training session using virtual reality technology is shown in Figure 6.

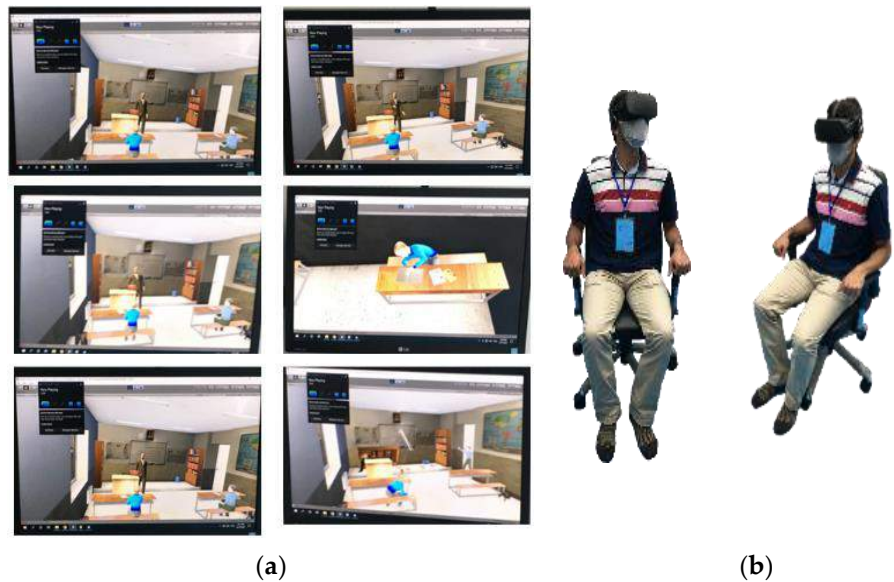

(a) (b)

**Figure 6.** Earthquake training in a virtual environment (**a**) Virtual scenario, (**b**) Experiment apparatus setup.

## 4. Results

### 4.1. Evaluating the Impact of VR on Safety Training Quality

Two weeks after the initial training sessions, both groups took a 10-min assessment test based on the questions listed in Table 1. The two groups' results have been provided in Tables A1 and A2 in the Appendix section.

When comparing the answers of the first and second groups, the second group's answers were more accurate and corresponded significantly with the first and second scenarios in VR. Some incorrect responses can be viewed in Table A2 in the Appendix section as participants in the first group drew on outside knowledge. For example, the first question asks about the safest place in a classroom during an earthquake. All the students in the second group answered that it was under a desk or a bench, but only two students in the first group gave the same answer. The remainder of the students responded with answers like "the door frame of the entrance" or "the schoolyard", both of which are incorrect. For the rest of the questions, the students in the second group gave accurate behavioral responses such as "staying calm and quiet" and named other safe locations in a classroom like "the corner of the classroom". In contrast, some of the students in the first group left their questions with no response. After this analysis, it can be concluded that the second group conducted the test more thoroughly and accurately.

*4.2. Evaluating How Readiness Affects an Individual's Anxiety and Decision-Making during an Earthquake*

To investigate the effect of readiness on an individual's anxiety and decision-making during earthquakes, the second group was exposed to two different scenarios within the VR model. Five students knew of the earthquake in the first scenario, and the other five were not informed. Hence, they assumed that it would be a typical educational setting in VR.

The DASS and BAI tests were taken before and after the experiment and were retaken two weeks after the initial date of the training. Table 2 and Table 3 list the results of the DASS tests before and after both scenarios.

**Table 2.** The results of the DASS test before the experiment.

| Sample | Depression | Grading | Anxiety | Grading | Stress | Grading |
|---|---|---|---|---|---|---|
| Sample 1 | 29 | Natural | 24 | Medium | 29 | Weak |
| Sample 2 | 7 | Natural | 2 | Natural | 19 | Natural |
| Sample 3 | 10 | Natural | 7 | Natural | 24 | Weak |
| Sample 4 | 19 | Natural | 14 | Natural | 26 | Weak |
| Sample 5 | 33 | Natural | 12 | Natural | 29 | Weak |
| Sample 6 | 29 | Natural | 17 | Weak | 30 | Weak |
| Sample 7 | 19 | Natural | 24 | Medium | 24 | Weak |
| Sample 8 | 36 | Weak | 14 | Natural | 13 | Natural |
| Sample 9 | 26 | Natural | 2 | Natural | 36 | Medium |
| Sample 10 | 17 | Natural | 10 | Natural | 19 | Natural |

**Table 3.** The results of the DASS test after the experiment.

| Sample | Depression | Grading | Anxiety | Grading | Stress | Grading |
|---|---|---|---|---|---|---|
| Sample 1 | 31 | Natural | 35 | Severe | 45 | Medium |
| Sample 2 | 12 | Natural | 16 | Natural | 32 | Medium |
| Sample 3 | 8 | Natural | 26 | Medium | 49 | Severe |
| Sample 4 | 17 | Natural | 30 | Medium | 40 | Medium |
| Sample 5 | 26 | Natural | 30 | Medium | 53 | Severe |
| Sample 6 | 32 | Natural | 25 | Medium | 53 | Severe |
| Sample 7 | 10 | Natural | 36 | Severe | 42 | Medium |
| Sample 8 | 32 | Natural | 19 | Weak | 53 | Severe |
| Sample 9 | 35 | Weak | 35 | Severe | 63 | Severe |
| Sample 10 | 11 | Natural | 26 | Medium | 47 | Medium |

Figure 7 is a graph illustrating Table 2 and Table 3 more precisely. The category of depression and stress was not considered to have a validation between anxiety scores between DASS and BAI. On the left side, the results of subsection 1 (the five participants who were aware of earthquake occurrence) have been illustrated before and after the experiment. Also, the right-side graph demonstrates the anxiety scores for subsection 2 (the five participants who were unaware of earthquake occurrence) before and after the experiment.

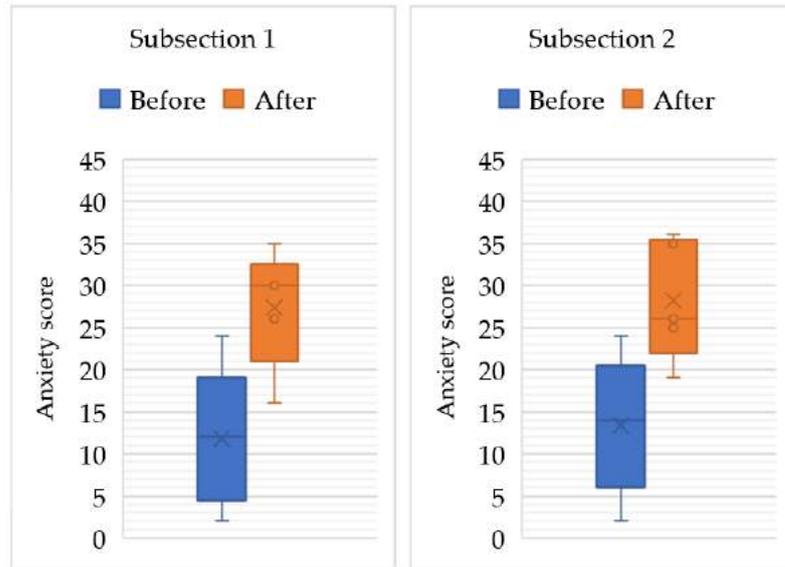

**Figure 7.** Levels of anxiety in the DASS test.

Not surprisingly, the subsection that was informed of the earthquake in the first scenario experienced less of an increase in their anxiety levels than the subsection that was uninformed of the earthquake. For the sake of concision, the second subsection will refer to the five uninformed students. In Figure 7, there is a higher level of incrementation among the second subsection relative to the first. The average increase in the first and second subsections is 18.4 and 27.2, respectively.

In Figure 8, the natural, weak, medium, and severe grades correspond to the colors blue, orange, gray, and yellow, respectively. Before the experiment, four students had the natural grade, one student had the medium grade in subsection 1, four students had a natural grade, and one student had the weak grade in subsection 2. After the experiment, the second subsection had two students in the medium grade and two students in the severe grade. While the grading had increased for all students, the increase was considerably more remarkable in the second subsection than in the first.

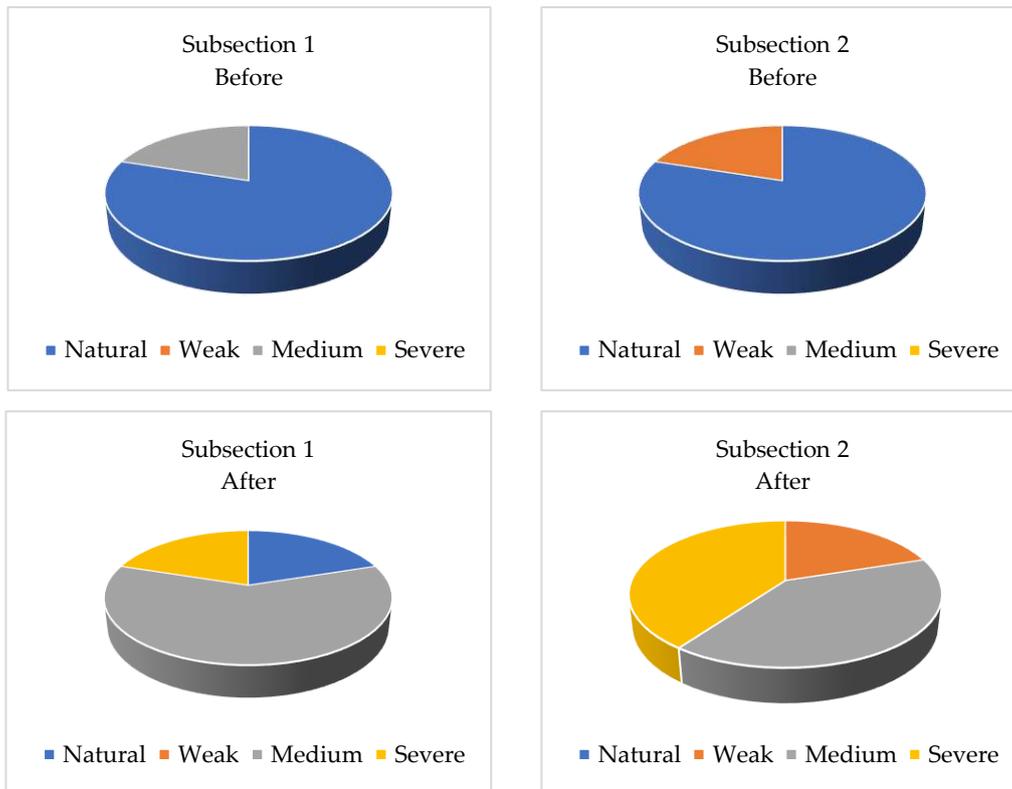

**Figure 8.** Grade of anxiety in the DASS test.

The anxiety levels of the second group and its subsections in the BAI test are presented in Table 4 and depicted in Figure 9. The results of the BAI are similar to the DASS test: both of these tests report an increase in all cases and a higher increase for the second subsection. The difference shows how the ability to foretell an earthquake can decrease the anxiety levels in individuals during an earthquake. Before the experiment, the grades for the first subsection were "Minimal Range" for three students and "Mild" for two students. After the experiment, these grades changed to three in "Moderate," one in "Mild," and one in "Severe" for the first subsection. The second subsection scored three grades of "Minimal Range" and two grades of "Mild" before the experiment and increased to three grades of "Severe" and two grades of "Moderate" after the experiment. The average increase for the first and second groups is equivalent to 13 and 21.8, respectively. Simply stated, this means that the second subsection experienced a more significant increase than the first subsection after the experiment.

**Table 4.** The results of the BAI test.

|  | Before | | After | |
| --- | --- | --- | --- | --- |
| Sample | Score | Garde | Score | Grade |
| Sample 1 | 12 | mild | 17 | moderate |
| Sample 2 | 3 | minimal range | 14 | mild |
| Sample 3 | 3 | minimal range | 19 | moderate |
| Sample 4 | 5 | minimal range | 22 | moderate |
| Sample 5 | 14 | mild | 30 | severe |
| Sample 6 | 4 | minimal range | 20 | moderate |
| Sample 7 | 8 | mild | 37 | severe |
| Sample 8 | 6 | minimal range | 29 | severe |
| Sample 9 | 12 | mild | 40 | severe |
| Sample 10 | 5 | minimal range | 18 | moderate |

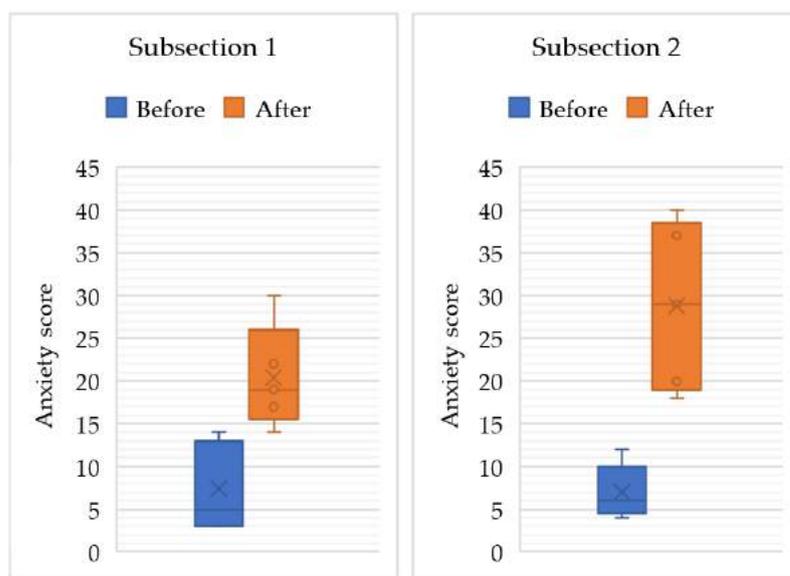

**Figure 9.** Box chart of anxiety scores in the BAI test.

Figure 10 describes the grade of anxiety levels from the BAI test in the second group. The graph can also be interpreted as a more significant increase in anxiety levels for the second subsection.

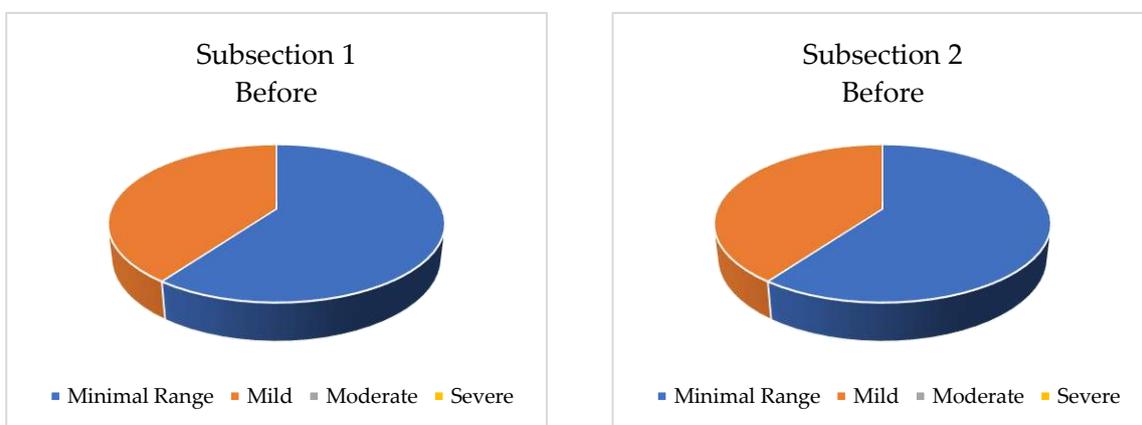

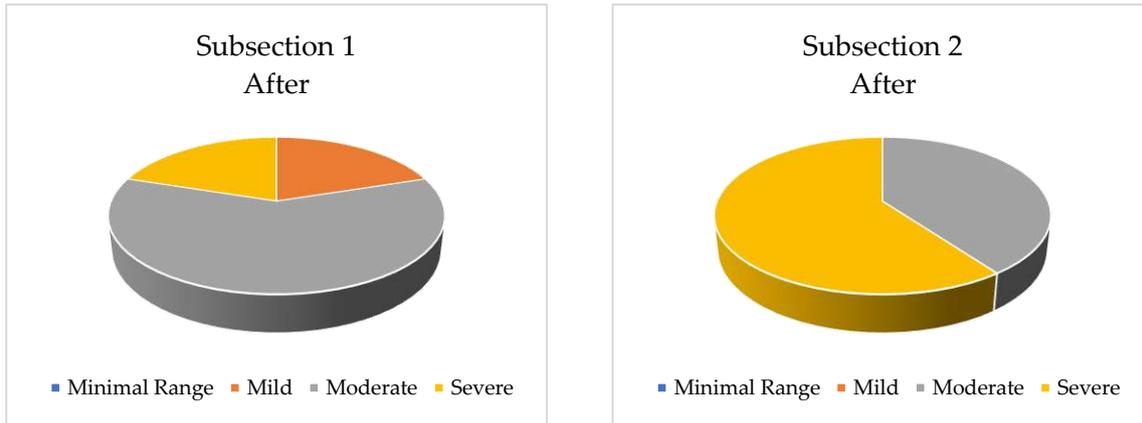

**Figure 10.** Grade of anxiety in the BAI test.

**5. Discussion**

According to recent studies in cognitive science (especially about anxiety, stress and decision-making), stress is found at the behavioral and neural levels. In other words, regions of the brain that regulate reasoning and decision-making are influenced by changes in stress and anxiety, and it can also change situational awareness [59]. These researchers have reached a common consensus that individuals who are stressed or in stressful situations will make more harmful and impulsive decisions. A severe increment in an individual's stress level will negatively impact his or her ability to rationalize and make proper decisions [59–61], which can often be seen in circumstances with high risk or reward [59,60]. As examined in this study, VR technology provides a basis for evaluating how premonition affects anxiety and decision-making.

Furthermore, this study explores the most successful ways to relay safety measures against earthquakes and the remarkable capabilities of VR technology. To summarize, it has been revealed that educating others through VR is much more effective than the in-person method of education and that individuals who are unaware of an earthquake occurring beforehand are proportionally more stressed than individuals who are aware. Alert systems that notify people of earthquakes in advance have been implemented in contemporary countries like Japan. This can minimize stress and anxiety during an earthquake and govern the decisions people will make in preparing for one.

The results of this study can be summarized as follows:

- Using VR technology and applying a virtual environment to educate safety measures against earthquakes can be more effective teaching than the in-person way in this field.
- By simulating the conditions of an earthquake in VR, people can experience a rare natural disaster like a large magnitude earthquake. This will assist them in regulating their anxiety and behaviors in the event of an actual earthquake.
- Installing alarms throughout a city and implementing a national alert system can increase the probability that people will act swiftly and safely in response to an earthquake.
- VR imitates real life through its sensory feedback and immersive environment. The second group that participated in the VR scenarios experienced a general increase in anxiety, which proves that VR is more than capable of creating realistic scenarios.

*5.1. Limitations of the Research*

Because this study was conducted during the COVID-19 pandemic, only a certain number of participants could be recruited to follow guidelines pertaining to the research. If the number of samples increases in a future study, it should hypothetically lead to a more precise and comprehensive conclusion. Moreover, the lack of an earthquake-shaking table and the confined spaces of the research laboratory contributed to the difficulties of creating a more realistic environment. Should a shaking table be present in a future study, it may assist in obtaining more extensive and accurate results.

*5.2. Suggestions for Future Research*

Because not many studies are available on the topic of earthquake-related safety measures, there are diverse paths researchers can take to broaden the scope of this field. Some suggested research topics are listed below:

- Evaluating how age, gender, and education level can affect the ability to learn and make decisions regarding earthquakes and earthquake-related safety measures
- Educating how to escape fire breakouts after an earthquake by using VR simulations
- Investigating decision-making and stress management processes during an earthquake by using brain and heart tests like EEG and HRV
- Examining other technological advancements like augmented reality as a way to educate safety measures against earthquakes

**6. Conclusions**

In recent years, the importance of safety measures against earthquakes has increased significantly. Earthquake-prone countries have been educating the public on the appropriate measures against earthquakes. It should be the duty of governments, businesses, and policymakers to assist in training current and future generations on how to safely navigate natural and manmade disasters like floods, earthquakes, fires, etc. Although Iran has implemented educational programs in offices and schools to raise public awareness, it has failed to impact Iranian society. These programs are widely ignored, which has regrettably led to financial damages and casualties when earthquakes occur. While structural engineers have written various guidelines on constructing structurally resistant buildings, governments and policymakers have not considered how educating safety measures the public could potentially save lives during an earthquake. Posters, movies, and conferences on this matter are small steps towards increasing awareness, but not introducing people to an earthquake's simulations will leave them mentally and emotionally unprepared when encountering a real one. Consequently, people will be more likely to make rash and harmful decisions due to unprecedented stress levels.

VR receives attention from researchers in many fields and has already been used in other seismic research studies. VR technology enhances an individual's senses, creating realistic scenarios when simulating earthquakes. The knowledge gained from experiencing an earthquake in VR can assist a person in making the right choices and managing their stress in the case of a real earthquake. In being aware of safety measures against earthquakes, individuals have the power to control their stress and make the best decisions for themselves.

In conclusion, this study has evaluated the effect of education and foresight on rational decisions during an earthquake. To enact this experiment, VR simulations were introduced to realistically model earthquakes, and self-reporting assessment tests were conducted to assess the comprehension levels after learning about earthquake-related safety measures. It can be reasonably inferred that VR is a successful educational tool and should be distributed on a broader scale for earthquake training purposes. Finally, predicting large earthquakes and implementing digital and/or physical national alert systems can significantly help to guide people's decision-making processes during an earthquake.

**Author Contributions:** Conceptualization, M.S.R.; methodology, M.S.R., H.T. and S.M.Z; software, M.S.R.; validation, M.S.R. and S.M.Z; formal analysis, M.S.R.; investigation, M.S.R.; resources, M.S.R. and H.T.; data curation, M.S.R.; writing—original draft preparation, M.S.R.; writing—review and editing, M.S.R., H.T. and S.M.Z; visualization, M.S.R; supervision, H.T. and S.M.Z.; project administration, M.S.R; funding acquisition, M.S.R. All authors have read and agreed to the published version of the manuscript.

**Funding:** This research received no external funding.

**Institutional Review Board Statement:** Ethical review and approval were waived for this study due to the absence of Personal Identifiable Information (PII) in the data collected by a research team member. Informed Consent Statement: Informed consent was obtained from all subjects involved in the study.

**Data Availability Statement:** The data presented in this study are available on request from the corresponding author. The data are not publicly available due to privacy issues.

**Acknowledgments:** The authors would like to acknowledge TECNOSA R&D Center and National Brain Mapping Laboratory for their sincere support. They are also thankful to Reza Rostami, Mojtaba Noghabaei, Neda Mohammadi, Mohammadali Keyvan, and Emmy Ly, and those who participated in this study and supported the research project. Also, the authors would like to acknowledge VT's OASF support for covering the fees of publishing this article. This study was presented as a Preprint on Arxiv [62].

**Conflicts of Interest:** The authors declare no conflict of interest.

**Appendix**

Table A1. The results of the assessment test for the first group.

|  | Question 1 | Question 2 | Question 3 | Question 4 | Question 5 |
|---|---|---|---|---|---|
| Participant 1 | The frame of the entrance door | Under the desk, the frame of the entrance door | Beside window | Keeping distance from fragile and sharp objects | I do not know |
| Participant 2 | In yard and street | Under the strong desk and door's frame | Under luster or ceiling fan or glass table | Escape | Keep a safe distance from buildings or tall things |
| Participant 3 | Under desk | Door's frame, corner of the class | Besides window, under the luster, beside the bookshelf, under anything that might fall | Staying calm and finding a safe location | We should stand in an open space where something is not present to fall |
| Participant 4 | Under desk | Far from the window, under the door's frame, in the same location but hand overhead | Besides windows, close to large equipment such as fan, air conditioner, … | Staying calm and keeping distance from dangerous items | Staying without movement and not getting close to the schoolyard |
| Participant 5 | Door's frame | Under desks and chairs | Besides wall and board, staying under a roof without shelter | Taking shelter in a safe location, we should keep a distance from areas that might be dangerous due to earthquakes, such as chemical materials or…. | Keep distance from walls and buildings, we should go to the center to keep distance from walls or things that might collapse |
| Participant 6 | Under desks | Door's frame | Besides windows and glasses, beside the collapsing wall | We should take shelter under desks or doors frames | We should keep our distance from falling items or go under desks to take shelter |
| Participant 7 | Underframe, which is made of | Desk, door's frame | Beside a window or under the open roof | Harnessing equipment and objects that might topple or fall | We should sit in the yard in an open space |
| | Iron and has high strength against breaking | | | | |
| Participant 8 | Door frame | Under benches and hold the desk columns- corner of the wall that does not have a window | Close to window, beside bookshelf that might fall down | We should keep a safe distance from shelves, so they do not fall on us during the earthquake, or other accident does not occur | We should keep our distance from the surroundings buildings |
| Participant 9 | Door frame, corner of walls | Under desks | Close to windows, closets, under cooler | Checking gas and water pipes, dangerous chemical materials | Keeping distance from the building |
| Participant 10 | Door frame, under desks by keeping distance from windows | Corners that are distant from windows | Under the windows, ceiling fan, and possible lusters | Keeping distance from shelves and dangerous items | Keeping distance from buildings and getting closer to the middle of the yard |

Table A2. The results of the assessment test for the second group.

| | Question 1 | Question 2 | Question 3 | Question 4 | Question 5 |
|---|---|---|---|---|---|
| Participant 1 | Under desk | Under door's frame, without glass and triangular corners of objects, under desks | Besides windows, under the fan, under boards, or painting board | If there is a flaming gas, it should be turned off quickly and then go under the desks | We should go to the middle of the yard with the most distance from tall objects and building in all directions |
| Participant 2 | Under desk | Door's frame, corner of the wall | Close to a window or any place that something might fall (Projector, cooler) | Keeping distance from objects that might fall and maintaining distance from explosive or acidic materials in the laboratory | Keeping distance from buildings that do not get trapped under it in case of collapse |
| Participant 3 | Under teachers' and students' desk | Under the desk of each person- door's frame-junction of walls- laying down on the floor with hands overhead | Under fluorescent light bulbs, besides window | Glasses should be covered by tape so they do not turn into broken parts | We should keep our distance from unsafe structures with high elevation |
| Participant 4 | Taking shelter under the desk and covering head by hand | Staying under door's frame, staying beside desk or big chair and covering head by hands while sitting, keeping distance from windows and closets | Close to window, glasses, under projector and items that are attached to wall or ceiling, close to closets | Taking shelter under a desk, covering your head with your hands, taking shelter under a door's frame while you are far from dangerous factors | Moving toward the middle of the yard, where there is no tall building or heavy object close to it |

| Participant | | | | | |
|---|---|---|---|---|---|
| Participant 5 | Under the classroom's desk | Corner of walls | Besides the window, board, closet, and ceiling light- the center of the room | Holding bag over your head and neck and sitting under a desk - keeping distance from dangerous items and encouraging others to keep calm and follow safety regulations | Keeping distance from structure and walls and wells location |
| Participant 6 | Under desk | Under the chair and door's frame | Under the window, picture frame, under the bookshelf, and fan | Going under desk | Keep distance from building and cover head and neck with hand |
| Participant 7 | Under desk | Far from windows and falling and fragile objects – close to main columns | Close to fragile and falling objects | Under the desk, or exiting the location if possible | Keeping distance from the building |
| Participant 8 | Under desk | Corner of walls | Door's frame, middle of the class, bookshelf | Staying calm and taking safe shelter | Keeping distance from surrounding building |
| Participant 9 | Under desk | Corner of main walls | Keeping distance from the bookshelf, door's frame, windows, glasses, fragile objects | Taking shelter under an empty desk without dangerous items | Staying in the center of the yard, far from the building |
| Participant 10 | Under desk | Keeping distance from the center of class- staying close to walls, corners far from the window | In front of the bookshelf and other falling objects and center of the class under the fluorescent light bulb | Keeping distance from dangerous objects and searching for shelter | Increasing distance from the school's building |